\documentclass[aps,prl,nofootinbib,superscriptaddress,reprint]{revtex4-1}
\usepackage{graphicx}
\usepackage{amsmath}

\begin{document}
\newcommand{\comment}[1]{\textit{ comment: #1}}
%\linenumbers

\title{Search for a Dark Matter annihilation signal from the Galactic Center halo with H.E.S.S.}
\date{\today}

\author{A.~Abramowski} 
 \affiliation{Universit\"at Hamburg, Institut f\"ur Experimentalphysik, Luruper Chaussee 149, D 22761 Hamburg, Germany}
 \author{F.~Acero} 
 \affiliation{Laboratoire de Physique Th\'eorique et Astroparticules, Universit\'e Montpellier 2, CNRS/IN2P3, CC 70, Place Eug\`ene Bataillon, F-34095 Montpellier Cedex 5, France}
 \author{F. Aharonian}  
 \affiliation{Max-Planck-Institut f\"ur Kernphysik, P.O. Box 103980, D 69029 Heidelberg, Germany}
 \affiliation{Dublin Institute for Advanced Studies, 31 Fitzwilliam Place, Dublin 2, Ireland}
 \affiliation{National Academy of Sciences of the Republic of Armenia, Yerevan}
 \author{A.G.~Akhperjanian} 
 \affiliation{Yerevan Physics Institute, 2 Alikhanian Brothers St., 375036 Yerevan, Armenia}
 \affiliation{National Academy of Sciences of the Republic of Armenia, Yerevan}
 \author{G.~Anton} 
 \affiliation{Universit\"at Erlangen-N\"urnberg, Physikalisches Institut, Erwin-Rommel-Str. 1, D 91058 Erlangen, Germany}
 \author{A.~Barnacka}
 \affiliation{Nicolaus Copernicus Astronomical Center, ul. Bartycka 18, 00-716 Warsaw, Poland}
 \affiliation{CEA Saclay, DSM/IRFU, F-91191 Gif-Sur-Yvette Cedex, France}
 \author{U.~Barres de Almeida} 
 \thanks{supported by CAPES Foundation, Ministry of Education of Brazil}
 \affiliation{University of Durham, Department of Physics, South Road, Durham DH1 3LE, U.K.} 
 \author{A.R.~Bazer-Bachi}
 \affiliation{center d'Etude Spatiale des Rayonnements, CNRS/UPS, 9 av. du Colonel Roche, BP 4346, F-31029 Toulouse Cedex 4, France}
 \author{Y.~Becherini}
 \affiliation{Astroparticule et Cosmologie (APC), CNRS, Universit\'{e} Paris 7 Denis Diderot, 10, rue Alice Domon et L\'{e}onie Duquet, F-75205 Paris Cedex 13, France}
 \affiliation{Laboratoire Leprince-Ringuet, Ecole Polytechnique, CNRS/IN2P3, F-91128 Palaiseau, France}
 \thanks{UMR 7164 (CNRS, Universit\'e Paris VII, CEA, Observatoire de Paris)}
 \author{J.~Becker}
 \affiliation{Institut f\"ur Theoretische Physik, Lehrstuhl IV: Weltraum und Astrophysik, Ruhr-Universit\"at Bochum, D 44780 Bochum, Germany}
% ...

\collaboration{H.E.S.S. Collaboration}\noaffiliation

\begin{abstract}
A search for a very-high-energy (VHE; $\geq 100$~GeV) $\gamma$-ray signal from self-annihilating particle Dark Matter (DM) is performed towards a region of projected distance $r\sim 45-150$~pc from the Galactic Center. The background-subtracted $\gamma$-ray spectrum measured with the  High Energy Stereoscopic System (H.E.S.S.) $\gamma$-ray instrument in the energy range between 300~GeV and 30~TeV shows no hint of a residual $\gamma$-ray flux. Assuming conventional Navarro-Frenk-White (NFW) and Einasto density profiles, limits are derived on the velocity-weighted annihilation cross section $\langle\sigma v\rangle$ as a function of the DM particle mass. These are among the best reported so far for this energy range. In particular, for the DM particle mass of $\sim1$ TeV, values for $\langle\sigma v\rangle$ above $3 \times 10^{-25} \text{ cm}^{3}\text{ s}^{-1}$ are excluded for the Einasto density profile. The limits derived here  differ much less for the chosen density profile parametrizations, as opposed to limits from $\gamma$-ray observations of dwarf galaxies or the very center of the Milky Way, where the discrepancy is significantly larger. 

%precise knowledge of the shape of the DM density profile at the center is lacking
%do not depend strongly on the profile
%% 
%% As ,  especially if an Einasto-type DM profile is realized by nature.
\end{abstract}

\maketitle
\section{Introduction}

The existence of particle DM is a widely accepted astrophysical concept that is used to explain, e.g., the formation of large scale structure during the evolution of the universe. In the present universe, galaxies, among others, are believed to be embedded in DM halos, which have density profiles that are much more extended than the profiles of visible matter and also exhibit the highest DM density in their centers (e.g. \cite{NFW:1996,Springel:2008,Diemand:2008}). This fact makes galactic centers promising targets to search for signals from annihilation or decay of hypothetical DM particles.  In simulations of cosmological structure formation, cold DM, which, among others, might consist of non-baryonic, weakly interacting massive particles ($m_{\mathrm{\chi}} \sim$ 10~GeV up to a few TeV) \cite{Griest:1988}, is able to reproduce the observed large scale structure of the universe \cite{Springel:2008,Diemand:2008}. They are expected to either annihilate or decay into standard model particles, producing, among other particles, photons in the final state. These photons are predicted to exhibit a continuous spectrum and to cover a broad energy range up to the DM mass, with possibly narrow lines from direct $\chi\chi\rightarrow\gamma\gamma / \gamma Z$ annihilations or other spectral features overlaid (see, e.g., \cite{Srednicki:1986, Bergstroem:1988, Bergstroem:2007}). Therefore, $\gamma$-ray observations, in particular at very high energies where DM masses of a few 100~GeV and above are probed, are a promising way to detect DM in space and to constrain its particle physics properties.
%Such particles are predicted, e.g., in various extensions of the standard model (SM) of particle physics (see, e.g., \cite{Bertone:2005} and references therein)}

A successful DM observation strategy has to avoid sky regions with strong astrophysical $\gamma$-ray signals, and should focus at the same time on regions with an expectedly large DM density. Besides dwarf galaxies, from which limits on $\langle\sigma v\rangle$ have been derived recently from observations at high (100~MeV-20~GeV) \cite{Scott:2010,Abdo:2010} and very high energies \cite{Aharonian:2008dm,Wood:2008,Aharonian:2010dmerr,Albert:2008,Aharonian:2009dm,Veritas:2010dm}, the Galactic Center (GC) region is a prime target for DM search, both because of its proximity and its predicted large DM concentration (see, e.g., \cite{NFW:1996}). As opposed to dwarf galaxies, however, the search for DM induced $\gamma$-rays in the GC is hampered by a strong astrophysical background. Especially at the very center, there is the compact $\gamma$-ray source HESS~J1745-290, coincident with the position of the supermassive black hole Sgr~A* and a nearby pulsar wind nebula \cite{Aharonian:2004wa,GCSpectrum,vanEldik:2010}, which makes the detection of $\gamma$-rays from DM annihilation difficult \cite{Aharonian:2006wh}. Furthermore, there is a band of diffuse emission along the GC ridge \cite{Aharonian:2006au}, which has already been used to derive upper limits on a DM induced signal \cite{Crocker:2010, Bertone:2009, Meade:2010}.

\begin{figure}
	\includegraphics[width=0.5\textwidth]{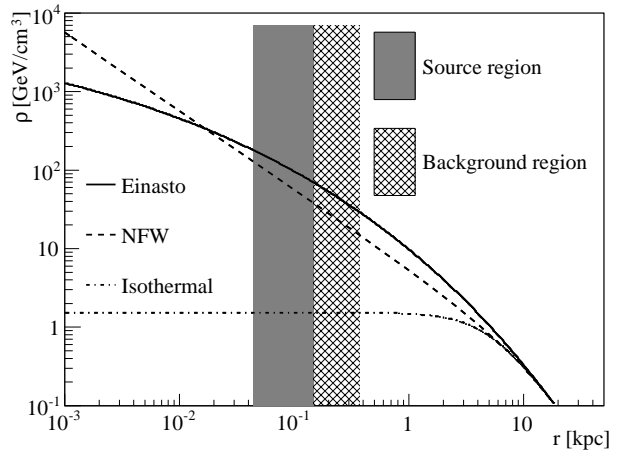}
	\caption{Comparison of the Galactic DM halo profiles used in this analysis. The parameters for the NFW and Einasto profiles are taken from \cite{Pieri:2009}. An isothermal profile \citep{Bertone:2005}, exhibiting a flat DM density out to a galactocentric distance of ~1~kpc, is shown for comparison. All profiles are normalized to the local DM density ($\rho_0 = 0.39$~GeV/cm$^{3}$ \cite{Catena:2009} at a distance of 8.5~kpc from the GC). The source region and the region used for background estimation are indicated. Note that the predicted DM density is always larger in the source region, except for the isothermal profile, which is included for completeness.}
	\label{Fig:Profiles}
\end{figure}

The $\gamma$-ray flux expected from annihilations of DM particles of mass $m_\chi$ is given by the product of the integral $J$ of the squared DM mass density $\rho$ along the line-of-sight, the velocity-weighted annihilation cross section $\langle\sigma v\rangle$, and the average $\gamma$-ray spectrum $dN/dE_\gamma$ produced in a single annihilation event \cite{Bertone:2005}. 
For a given DM mass and given branching ratios for annihilation into SM particles, the shape of the expected $\gamma$-ray spectrum, determined by $dN/dE_{\gamma}$, can be rather accurately computed, since only SM particle physics processes are involved. Measuring the velocity-weighted annihilation cross section $\langle\sigma v\rangle$, however, suffers from large systematic uncertainties on the astrophysical factor $J$. This is because in most indirect searches the main contribution of the total DM signal arises from the very central region of the object under study, where the DM density profile peaks. In these inner regions, however, the profile is so far only poorly known. In particular, for a Milky-Way sized DM halo, the radial DM density profiles obtained by the Aquarius \citep{Springel:2008} and Via Lactea II \citep{Diemand:2008} simulations can be described by Einasto and NFW parametrizations, respectively \citep{Pieri:2009}. These profiles are shown in Fig.~\ref{Fig:Profiles} as a function of galactocentric distance $r$. Large differences between both parametrization s occur if they are extrapolated down to the very center of the halo, where the NFW profile is more strongly peaked. At distances $>10$~pc, however, the difference is merely a factor of two, allowing one to put limits on $\langle\sigma v\rangle$ which do not depend strongly between either of the two parametrizations.
%are observed at 
%depend strongly on the still poorly constrained DM halo properties

Here we exploit this fact by searching for a VHE $\gamma$-ray signal from DM annihilation in our own Galaxy, in a region with a projected galactocentric distance of $45$~pc~$-150$~pc (corresponding to an angular distance of $0.3^\circ - 1.0^\circ$)\footnote{Here and in the following a distance of the GC to the observer of 8.5~kpc is assumed.}, excluding the Galactic plane. In this way, contamination from $\gamma$-ray sources in the region is naturally avoided as well.

\section{Methodology\label{sec:Methodology}}

The analysis is carried out using 112~h (live time) of GC observations with the H.E.S.S. VHE $\gamma$-ray instrument (see \cite{Crab:2006} and references therein) taken during the years 2004-2008. For minimum energy threshold, only observations with zenith angles smaller than $30^\circ$ are considered. The mean zenith angle is $14^\circ$.  To avoid possible systematic effects, pointing positions were chosen fairly symmetric with regard to the Galactic plane. The mean distance between the pointing position and the GC is $0.7^\circ$, with a maximum of $1.5^circ$. Events passing H.E.S.S. standard cuts defined in \cite{Crab:2006} are selected for analysis. To minimize systematic uncertainties due to reduced $\gamma$-ray efficiency at the edges of the $\sim 5^\circ$ diameter field-of-view (FoV) of H.E.S.S., only events reconstructed within the central $4^\circ$ are considered.  The effective $\gamma$-ray collection area of the chosen event selection is $\approx 1.7 \times 10^{5} \text{ m}^{2}$ at 1 TeV. The total effective exposure of the utilized dataset at 1 TeV amounts to $\approx 2.4 \times 10^{7} \text{ m}^{2}$ sr s.

\begin{figure}
	\includegraphics[width=0.5\textwidth]{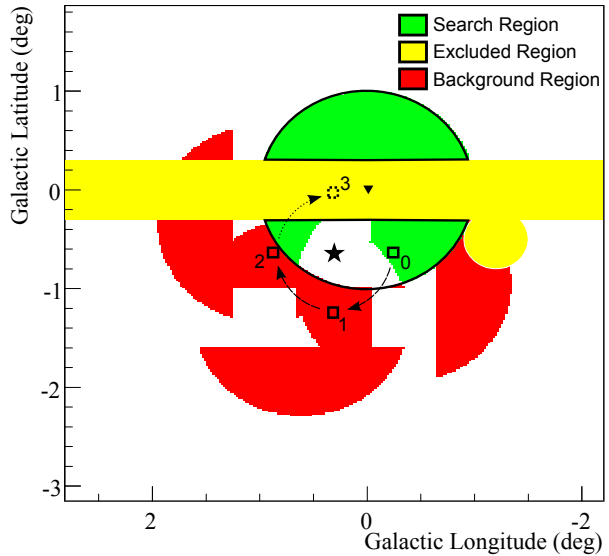}
	\caption{Illustration of the cosmic ray background subtraction technique for a single telescope pointing position (depicted by the star).  Note that this position is only one of the several different pointing positions of the dataset. The DM source region is the green area inside the black contours, centered on the GC (black triangle). Yellow regions are excluded from the analysis because of contamination by astrophysical sources. Corresponding areas for background estimation (red regions) are constructed by rotating individual pixels of size $0.02^\circ\times 0.02^\circ$ of the source region around the pointing position by $90^\circ$, $180^\circ$, and $270^\circ$. This choice guarantees similar $\gamma$-ray detection efficiency in both the source and background regions. As an example, pixels labeled \emph{1} and \emph{2} serve as background control regions for pixel \emph{0}. Pixel \emph{3} is not considered for background estimation because it  is located in an excluded region. Pixels in the source region, for which no background pixels can be constructed, are not considered in the analysis for this particular pointing position and are left blank.}
	\label{Fig:ReflectedPixel}
\end{figure}

$\gamma$-rays from DM annihilations are searched for in a circular \emph{source region} of radius $R_\mathrm{on}=1.0^\circ$ centered at the GC. Contamination of the DM signal by local astrophysical $\gamma$-ray sources is excluded by restricting the analysis to Galactic latitudes $|b|>0.3^\circ$, effectively cutting the source region into two segments above and below the Galactic plane (see Fig. \ref{Fig:ReflectedPixel}). Simulations show that both for Einasto and NFW parametrizations the sensitivity varies only within a few percent when varying the source region size in the range $0.8^\circ \leq R_\mathrm{on} \leq 1.2^\circ$. 
%\footnote{When enlarging the source region, the sensitivity increases due to the total enclosed astrophysical factor $J$, but decreases at the same time due to the growing number of collected misidentified cosmic ray background events, which are isotropically distributed.} 

For ground-based VHE $\gamma$-ray instruments like H.E.S.S., any $\gamma$-ray signal is accompanied by a sizable number of cosmic-ray induced background events, which are subtracted from the source region using control regions in the FoV of the observation. Fig.~\ref{Fig:ReflectedPixel} visualizes details of the method, which is an evolution of the standard \emph{reflected background} technique \cite{Berge:2006ae} adjusted for this particular analysis.
%\footnote{Most of the archival H.E.S.S. observations were pointed inside the source region. With the standard technique, this would lead to an overlap of source and background regions, rendering such observations unusable for the analysis presented.}
By construction, background regions are located further away from the GC than the source region. This is an important aspect, since, unavoidably, a certain amount of DM annihilation events would be recorded in the background regions, too, reducing a potential excess signal obtained in the source region. For the NFW and Einasto profiles, the expected DM annihilation flux is thus smaller in the background regions than in the source region (cf. Fig.~\ref{Fig:Profiles}), making the measurement of a residual annihilation flux possible. Note, however, that for an isothermal halo profile, the signal would be completely subtracted. As far as the background from Galactic diffuse emission is concerned, its predicted flux \cite{GDE:2005} is significantly below the current analysis sensitivity, thus its contribution is not further considered in the analysis. In any case, since its intensity is believed to drop as a function of Galactic latitude, $\gamma$-rays from Galactic diffuse emission would be part of a potential signal, and therefore lead to more conservative results for the upper limits derived in this analysis.
 
%coincides with an astrophysical $\gamma$-ray source

\section{Results}

\begin{figure}
\includegraphics[width=0.5\textwidth]{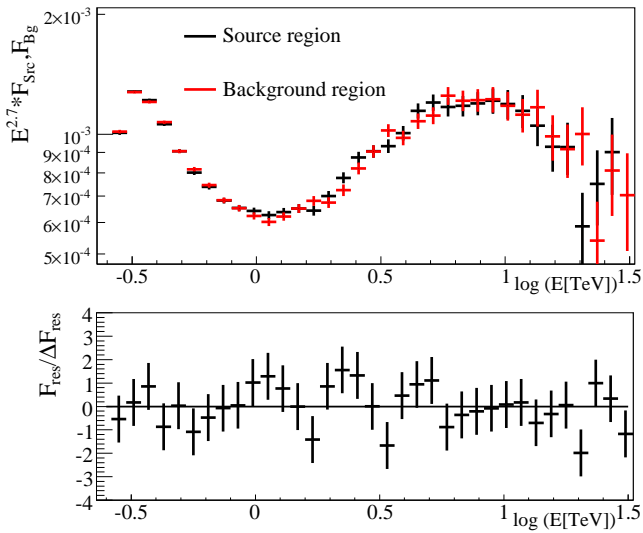}
\caption{Top panel: Reconstructed differential flux $F_{\text{Src/Bg}}$, weighted with $E^{2.7}$ for better visibility, obtained for the source and background regions as defined in the text. The units are TeV$^{1.7}$~m$^{-2}$~s$^{-1}$~sr$^{-1}$. Due to an energy-dependent selection efficiency and the use of effective areas obtained from $\gamma$-ray simulations, the reconstructed spectra are  modified compared to the cosmic-ray  power-law spectrum measured on Earth. Bottom panel: Flux residua $F_{\text{res}}/ \Delta F_{\text{res}}$, where $F_{\text{res}}=F_{\text{Src}}-F_{\text{Bg}}$ and $\Delta F_{\text{res}}$ is the statistical error on $F_{\text{res}}$. The residual flux is compatible with a null measurement. Comparable null residuals are obtained when varying the radius of the source region, subdividing the data set into different time periods or observation positions, or analyzing each half of the source region separately.}
\label{Fig:Spectra}
\end{figure}

Using zenith angle-, energy- and offset-dependent effective collection areas from $\gamma$-ray simulations, flux spectra shown in Fig. \ref{Fig:Spectra} are calculated from the number of events recorded in the source and background regions\footnote{The background spectrum is rescaled by the ratio of the areas covered by source and background regions (cf. also \cite{Berge:2006ae}).}. It should be stressed that these spectra consist of $\gamma$-ray-like cosmic-ray background events. Both source and background spectra agree well within the errors, resulting in a null measurement for a potential DM annihilation signal, from which upper limits on $\langle\sigma v\rangle$ can be determined.

%w with a spectral index of -2.7 (Fig. \ref{Fig:Spectra})

%, in units of $(8.5$~kpc$)^{-1}\cdot(0.3$~GeV/cm$^3)^{-2}$, 
The mean astrophysical factors $\bar{J}_\mathrm{src}$ and $\bar{J}_\mathrm{bg}$ are calculated for the source and background regions, respectively. The density profiles are normalized to the local DM density $\rho_0 = 0.39$~GeV/cm$^{3}$ \cite{Catena:2009}. Assuming an Einasto profile, $\bar{J}_\mathrm{src}=3142 \times \rho_\mathrm{E}^2 \times d_\mathrm{E}$ and $\bar{J}_\mathrm{bg}=1535 \times \rho_\mathrm{E}^2 \times d_\mathrm{E}$, where $\rho_\mathrm{E} = 0.3 \text{ GeV}/\text{cm}^3$ is the conventional value for the local DM density and $d_\mathrm{E} = 8.5$~kpc the distance of Earth to the GC. For a NFW profile, $\bar{J}_\mathrm{src}=1604 \times \rho_\mathrm{E}^2 \times d_\mathrm{E}$ and $\bar{J}_\mathrm{bg}=697 \times \rho_\mathrm{E}^2 \times d_\mathrm{E}$ are obtained. This means that for an assumed Einasto (NFW) profile, background subtraction reduces the excess DM annihilation flux in the source region by 49~\% (43~\%), which is taken into account in the upper limit calculation.

\begin{figure}
\includegraphics[width=0.5\textwidth]{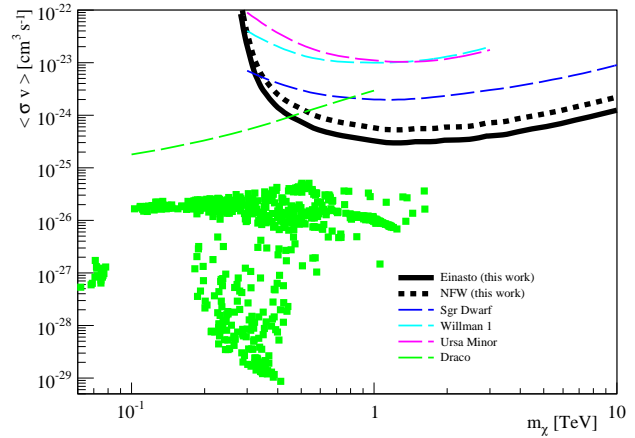}
\caption{Upper limits (at 95\% CL) on the velocity-weighted annihilation cross-section $\langle \sigma v \rangle$ as a function of the DM particle mass $m_{\chi}$ for the Einasto and NFW density profiles. The best sensitivity is achieved at $m_{\chi}\sim 1$~TeV.  %Limits, which are derived using the TMVA background rejection technique \cite{Ohm:2009}, are also shown for the Einasto profile. 
For comparison,  the best limits derived from observations of dwarf galaxies at very high energies, i.e. Sgr Dwarf \cite{Aharonian:2008dm}, Willman 1, Ursa Minor \cite{Veritas:2010dm} and Draco \cite{Abdo:2010}, using in all cases NFW shaped DM profiles, are shown. Similar to source region of the current analysis, dwarf galaxies are objects free of astrophysical background sources. The green points represent DarkSUSY models \cite{DarkSusy:2004}, which are in agreement with WMAP and collider constraints and were obtained with a random scan of the mSUGRA parameter space using the following parameter ranges: $10 \text{ GeV} < M_0 < 1000 \text{ GeV}$, $10 \text{ GeV} < M_{1/2} < 1000 \text{ GeV}$, $A_0 = 0$, $0 < \text{tan} \beta < 60$, $\text{sgn} (\mu) = \pm 1$.	}
\label{Fig:UpperLimits}
\end{figure}

Under the assumption that DM particles annihilate into quark-antiquark pairs and using a generic parametrization for a continuum spectrum of $\gamma$-rays created during the subsequent hadronization \cite{Tasitsiomi:2002, Hill:1987}, limits on $\langle \sigma v \rangle$ as a function of the DM particle mass are calculated for both density profiles (see Fig. \ref{Fig:UpperLimits}). These limits are among the most sensitive so far at very high energies, and in particular are the best for the Einasto density profile, for which at $\sim1$ TeV values for $\langle\sigma v\rangle$ above $3 \times 10^{-25} \text{ cm}^{3}\text{ s}^{-1}$ are excluded. As expected from the astrophysical factors, the limits for the Einasto profile are better by a factor of two compared to those for the NFW profile. Still, the current limits are one order of magnitude above the region of the parameter space where supersymmetric models provide a viable DM candidate (see Fig. \ref{Fig:UpperLimits}). Apart from the assumed density parametrizations and the shape of the $\gamma$-ray annihilation spectrum, the limits can shift by 30\% due to both the uncertainty on the absolute flux measurement \cite{Crab:2006} and the uncertainty of 15\% on the absolute energy scale. For the latter case, apart from a displacement with regard to the DM particle mass scale, the limits shift up (down) if the $\gamma$-ray energy is overall under(over)estimated.

%\footnote{Model-dependent effects like internal Bremsstrahlung or Sommerfeld enhancement are not considered in this work.} 
% of the DM annihilation products 
%The sensitivity of the measurement can be extended towards $m_{\chi}=200$ GeV by applying an improved background rejection method \cite{Ohm:2009}. 
%Compared to measurements using observations of the very central part of galaxies, including our own, where the uncertainties on the DM density can reach several orders of magnitude, the systematic uncertainty due to the unknown DM density profile is significantly reduced in this work.

\section{Summary}

A search for a VHE $\gamma$-ray signal from DM annihilations was conducted using H.E.S.S. data from the GC region. A circular region of radius $1^\circ$ centered at the GC was chosen for the search, and contamination by astrophysical $\gamma$-ray sources along the Galactic plane was excluded. An optimized background subtraction technique was developed and applied to extract the $\gamma$-ray spectrum from the source region. The analysis resulted in the determination of stringent upper limits on the velocity-weighted DM annihilation cross-section $\langle \sigma v \rangle$, being among the best so far at very high energies.  At the same time, the limits do not differ strongly between NFW and Einasto parametrizations of the DM density profile of the Milky-Way.
%The limits depend only weakly on the DM density profile shape, therefore the dependence of the results on astrophysical uncertainties is significantly reduced.

\begin{acknowledgments}
The support of the Namibian authorities and of the University of Namibia
in facilitating the construction and operation of H.E.S.S. is gratefully
acknowledged, as is the support by the German Ministry for Education and
Research (BMBF), the Max Planck Society, the French Ministry for Research,
the CNRS-IN2P3 and the Astroparticle Interdisciplinary Programme of the
CNRS, the U.K. Science and Technology Facilities Council (STFC),
the IPNP of the Charles University, the Polish Ministry of Science and 
Higher Education, the South African Department of
Science and Technology and National Research Foundation, and by the
University of Namibia. We appreciate the excellent work of the technical
support staff in Berlin, Durham, Hamburg, Heidelberg, Palaiseau, Paris,
Saclay, and in Namibia in the construction and operation of the
equipment.
\end{acknowledgments}

\bibliography{GCDMPaper}

\end{document}